\documentclass[12pt,preprint]{aastex}
\slugcomment{} \shorttitle{SDSS NLS1s} \shortauthors{Zhou et al.}

\begin{document}

\title{A Comprehensive Study of 2 000 Narrow Line Seyfert 1 Galaxies from
the Sloan Digital Sky Survey: I. The Sample}

\author{Hongyan Zhou\altaffilmark{1,3}, Tinggui Wang\altaffilmark{1,3}, Weimin Yuan\altaffilmark{2},
Honglin Lu\altaffilmark{1,3}, Xiaobo Dong\altaffilmark{1,3}, Junxian
Wang\altaffilmark{1,3}, and Youjun Lu\altaffilmark{1}}

\altaffiltext{1}{Center for Astrophysics, University of Science
and Technology of China, Hefei, Anhui, P.R.China}

\altaffiltext{2}{National Astronomical Observatories/Yunnan
Observatory, Chinese Academy of Sciences, Kunming, Yunnan, P.O. BOX
110, P.R.China}

\altaffiltext{3}{Joint Institute of Galaxies and Cosmology, SHAO and
USTC}

\email{mtzhou@ustc.edu.cn}

\begin{abstract}
This is the first paper in a series dedicated to the study of the
emission-line and continuum properties of narrow line Seyfert 1
galaxies (NLS1s). We carried out a systematic search for NLS1s from
objects assigned as ``QSOs'' or ``galaxies'' in the spectroscopic
sample of the Sloan Digital Sky Survey Data Release 3 (SDSS DR3) by
a careful modeling of their emission lines and continua. The  result
is a uniform sample comprising $\sim 2~000$ NLS1. This sample
dramatically increases the number of known  NLS1 by a factor of
$\sim 10$ over previous compilations. This paper presents the
parameters of the prominent emission lines and continua, which were
measured accurately with typical uncertainties $<10\%$. Taking
advantage of such an unprecedented large and uniform sample with
accurately measured spectral parameters, we carried out various
statistical analysis, some of which were only possible for the first
time. The main results found are as follows. (1) Within the overall
Seyfert 1 population, the incidence of NLS1s  is  strongly dependent
on the optical, X-ray, and radio luminosities as well as the
radio-loudness. The fraction of NLS1 peaks around SDSS g-band
absolute magnitude $M_{g}\sim-22^{m}$  in the optical and $\sim
10^{43.2}~erg~s^{-1}$ in the soft X-ray band, and decreases quickly
as the radio-loudness increases. (2) On average the relative FeII
emission, $R_{4570}=FeII(4434-4684)/H\beta$, in NLS1s is about twice
that in normal AGNs, and is anti-correlated with the broad component
width of the Balmer emission lines. (3) The well-known
anti-correlation between the width of broad low-ionization lines and
the soft X-ray spectral slope for broad line AGNs extends down to
$FWHM\sim 1~000~km~s^{-1}$ in NLS1s, but the trend appears to
reverse at still smaller line widths. (4) The equivalent width of
H$\beta$ and FeII emission lines are strongly correlated with the
H$\beta$ and continuum luminosities. (5) We do not find any
difference between NLS1s and normal AGNs in regard to the Narrow
Line Region. (6) We have examined the black hole mass vs.\ stellar
velocity dispersion ($M_{BH}-\sigma_{*}$) relation for a subsample
of 308 NLS1s for which $\sigma_{*}$ could be measured directly from
fitting the starlight in the SDSS spectra with our stellar spectral
templates. A significant correlation between $M_{BH}$ and
$\sigma_{*}$ is found, but with the bulk of  black hole masses
falling below the values expected from the $M_{BH}-\sigma_{*}$
relation for normal galaxies and normal AGNs. This result indicates
that NLS1s are underage AGNs, where the growth of the SMBH lags
behind the formation of the galactic bulge. (7) We also find that
the FWHM of [NII] line is well correlated with $\sigma_{*}$ in 206
NLS1s, for which both parameters could be derived with reasonable
accuracy. The [NII] width can predict the stellar velocity
dispersion to an accuracy of $\sim 30\%$. A similar
$M_{BH}-\sigma_{*}$ relation could be found for a larger sample of
613 NLS1s on making use of the more reliable measurements of
FWHM[NII].
\end{abstract}


\section{Introduction}
Active galactic nuclei (AGNs)---Seyfert galaxies and quasars---are
characterized by non-thermal continuum emission over almost the
whole electromagnetic spectrum from their nuclei and prominent
emission lines in the ultraviolet and the optical. The overall
spectral energy distribution (SED) can be described as an underlying
power law of the form $F_{\nu}\varpropto \nu^{-\alpha}$ interspersed
with bumps and dips. The AGNs can be classified into two types
according to their emission spectrum: type~1 AGNs  show both broad
emission lines with full width at half maximum (FWHM) of a few
thousand $km~s^{-1}$ and narrow emission lines of a few hundred
$km~s^{-1}$, while type~2 AGNs show only narrow emission lines.
Permitted and semi-forbidden lines are seen with both broad and
narrow profiles, while forbidden lines are  seen only with narrow
profiles. It was recognized at the end of 1970s (e.g., Osterbrock
1978) that such a split might be an orientation effect involving
obscuration of the broad-line region (BLR). According to unification
models, all AGNs have intrinsically the same physical structure, but
in type~2s the BLR and continuum emission region are obscured by a
presumed dusty torus located somewhere between the BLR and the
narrow line region (NLR).

It seems that the division of AGNs into type~1 and type~2 is
well-defined and the evidence for unification schemes is compelling
(cf, Antonucci 1993). Interestingly, almost at the same time when
the unified models were first brought forward, it was found that two
objects, Mrk 359 (Davidson \& Kinman 1978) and Mrk 42 (Koski 1978;
Phillips 1978), have all the spectral properties of a typical type~1
AGN, except that the ``broad permitted lines" are comparable in
width to the forbidden lines as in the case of type~2 AGNs. Some
other unusually narrow ``broad line emitters" (e.g., Mrk 493 and Mrk
783, Osterbrock \& Dahari 1983) were identified in the following
years. This intriguing mixture of features makes this ``peculiar"
group of AGNs very interesting. Osterbrock \& Pogge (1985, hereafter
OP85) denoted these objects as narrow line Seyfert 1 galaxies
(NLS1s). Their original classification included the following two
empirical and somewhat subjective criteria: 1) the Balmer lines are
only slightly broader than forbidden lines; and 2) the line ratio
$[OIII]\lambda5007/H\beta<3$; but exceptions are allowed when strong
high-ionization iron lines, such as [FeVII]$\lambda6087$ and
[FeX]$\lambda6375$ are present, which are seldom seen in type~2
AGNs. In his spectropolarimetric study of NLS1s, Goodrich (1989)
treated them as an extension of normal broad-line Seyfert~1s
(hereafter BLS1s) to lower FWHM of permitted lines and quantified
the first criterion of OP85 with $FWHM(H\beta)<2000~km~s^{-1}$.

Apart from the unusually narrow Balmer lines, NLS1s often show the
strong permitted FeII emission lines in their optical and
ultraviolet spectra; they also show some other extreme properties,
such as a steep soft X-ray spectrum, rapid X-ray variability, dense
line-emitting gas, and a commonly blue shifted UV line profile
(Boller et al.\ 1996; Leighly 1999; Aoki \& Yoshida 1999; Wills et
al.\ 1999; Leighly 2000). In fact, these properties are extensions
of the following correlations between various observables and the
$H\beta$ line width (Boroson \& Green 1992; Laor et al.\ 1994; Wang,
Brinkmann, \& Bergeron 1996; Xu et al. 2003): the narrower the $H\beta$ line, the
steeper the X-ray spectrum, the stronger the $H\beta$ blue wings,
the faster the X-ray variation, the stronger the optical FeII
emission, the weaker the [OIII] lines, and the denser the BLR
clouds. These form the so-called eigenvector 1 (E1, Boroson \& Green
1992; Sulentic et al. 2000). E1 is taken as the strongest and most
far-reaching type~1 AGN unification yet discovered (e.g., Marziani
et al.\ 2001), which incorporates the geometry, kinematics, and
physical condition of the central regions in AGNs. Its underlying
physical driver is related to either the basic  physical parameters
of the black hole accretion disk such as the black hole mass and
accretion rate, or geometrical parameters such as the viewing angle
(Boroson \& Green 1992; Boller et al.\ 1996; Wang, Brinkmann, \&
Bergeron 1996; Laor et al.\ 1997; Sulentic et al.\ 2000; Marziani et
al.\ 2001).  Since NLS1s lie at the extreme negative end of E1, they
may well occupy some extreme regions in the parameter space of some
primary physical parameters, whatever they are. We can, therefore,
use the unusualness of NLS1s to test the viability of the  different
AGN models.

However, some of the basic properties of NLS1s remain unclear
even though it has been two decades since their first identification.
We summarize these issues under the following  two headings.
\begin{description}
\item[Properties extensively explored, but no consensus reached yet:]
1) For Seyfert 1 galaxies and QSOs, a correlation between the soft
X-ray photon index (from ROSAT observations) and the $H\beta$ line
FWHM has now been well established (Laor et al.\ 1994; Boller et
al.\ 1996; Wang, Brinkmann, \& Bergeron 1996).
In comparison, for the NLS1s, which have on average steeper photon indices,
the same correlation shows a larger dispersion.
It is still not known whether the flatness of the X-ray spectra
of some NLS1s is intrinsic or is due to internal absorption.
2) Boroson \& Green (1992), McIntosh et al.\ (1999) and
Grupe et al.\ (1999) found
that the strength of the FeII and [OIII] lines are
anti-correlated, but this was not confirmed by V{\' e}ron-Cetty,
V{\' e}ron, \& Gon{\c c}alves (2001, hereafter VVG01). It is
unclear whether this correlation should be included in E1.
3) Rodr{\'{\i}}guez-Ardila et al.\ (2000) claimed that the emission
line ratio of the narrow line region (NLR), [OIII]/H$\beta$, is in
the range of $\sim 1-5$ for NLS1s, significantly less than the
values $\sim 10$, typical of normal AGNs, while VVG01 did  not
find any difference between the two types.
\item[Virgin territories:]
Compared to the well studied emission line and X-ray properties of
NLS1s, little or no attention has been given to 1) their
broad-band continuum properties; 2) the relationship between the
NLS1 phenomena and the evolution of the host galaxies; 3) the type~2
counterparts of NLS1s predicted by AGN unification models; and
4) their spectral variability.
\end{description}
To resolve the controversies and to seek new insights, it would be
helpful to have a large-size homogeneous sample of optically
selected NLS1s with accurately  measured parameters of both narrow
and broad lines. The Sloan Digital Sky Survey (SDSS, York et al.
2000) is expected to be able to provide such NLS1  samples resulting
from well-defined selection criteria that are large enough for
serious statistical studies; furthermore, it is expected that
potentially peculiar and interesting objects will be identified and
studied with follow-up observations. The first attempt in this
direction was made by Williams, Pogge, \& Mathur (2002, hereafter
WPM02) who compiled a sample of NLS1s from the SDSS Early Data
Release (EDR, Stoughton et al.\ 2002), which comprised 150 objects.
With a roughly  10-fold increase in size over the EDR, the presently
available SDSS spectroscopic data  can be expected to yield many
more NLS1s for the study.

We carried out a systematic search for NLS1s from the spectroscopic
sample of the Sloan Digital Sky Survey Data Release 3 (SDSS DR3,
Abazajian et al.\ 2005). In order to obtain from SDSS spectra
accurate and reliable line widths that are essential for the
selection and the follow-on analysis, we carefully fitted emission
lines, AGN continuum and starlight in a self-consistent manner. As a
result, we now have a uniform sample comprising as many as 2~011
NLS1s. The accurately measured emission-line parameters and hence
the selection of sample objects are an improvement on WPM02,
rendering our sample less affected by selection biases compared to
WPM02. This paper is the first of a series which reports on our
ongoing project for extensive studies of NLS1s by fully exploring
the potential of this large-size sample. Here we present the object
catalog, which will be kept updated as  new SDSS data are further
available; also presented are some preliminary statistical results
on the sample as a whole. We shall attempt to address some of the
issues raised above in forthcoming papers. The plan of this paper is
as follows: In the next section, we describe in detail the methods
we used to process all the low redshift QSOs and galaxies in the
SDSS DR3. The sample selection is described in \S3. The properties
of the sample are discussed in \S4, including a statistical analysis
of the emission lines and remarks on potentially interesting
individuals. We explore the black hole--bulge relation for NLS1s in
\S5. Our main results are summarized in the last section, followed
by prospects of future work. A $\Lambda$-dominated cosmology with
$H_0=70~km~ s^{-1}~Mpc^{-1}$, $\Omega_M=$ 0.3 and $\Omega_\Lambda=$
0.7 is assumed throughout the paper.

\section{Optical Spectral Analysis}

\subsection{An overview}
\label{sect:overview}

Our NLS1s were selected from the SDSS DR3 catalogs that are
classified as ``galaxy" or ``QSO" by the SDSS spectroscopic
pipeline. Only objects with a redshift $z \lesssim 0.8$ were
considered to ensure that the $H\beta$ and [OIII] emission lines
would fall within the wavelength coverage of the SDSS spectrograph
(Details of SDSS spectroscopy can be found in Stoughton et al.\
2002). The spectra of the resulting 387 483 extragalactic objects
(12 824 ``QSOs" and 374 659 ``galaxies") were first corrected for
the Galactic extinction using the extinction curve of Schlegel et
al. (1998), and then shifted back to their rest frames using the
redshifts provided by the SDSS pipeline.

{\em Contamination from stellar light:} The fibers that feed the
SDSS spectrograph subtend a diameter of $3^{''}$ on the sky, which
corresponds to $\sim 6.5~kpc$ at $z=0.1$. AGN spectra taken through
such a fixed aperture may include significant light from the host
galaxies, as exhibited in the composite spectra of SDSS quasars
(Vanden Berk et al.\ 2001). Careful removal of the starlight is
essential for reliable measurement of the emission lines, and hence
correct identifications of NLS1s from the  huge SDSS data. Actually,
the starlight ``contamination" provides valuable information about
the host galaxies, which is of interest in itself. For this and
other purposes,  we have developed a technique of properly modeling
the stellar component (see Lu et al.\ 2006 for details of the
method). This method decomposes the SDSS spectra of active galaxies
into stellar and non-stellar nuclear components, provided that  the
two components are comparable in strength. A description of this
method is given in  \S\,\ref{sect:starlight}.

{\em Broad-line and narrow-line decomposition:} The Balmer emission
lines of NLS1s may include significant contribution from narrow-line
components (VVG01; Rodr{\'{\i}}guez-Ardila et al.\ 2000). In the
selection of NLS1s based on the line's FWHM, it is therefore more
appropriate to use the {\em broad} component rather than the total
emission line, as was done in WPM02. Moreover, the de-blended narrow
components  have  interest of their own for the study of the NLR
properties in NLS1s. We therefore performed emission-line
decomposition in the analysis. For reliable line decomposition, good
spectral quality is required in terms of both S/N ratio and spectral
resolution. Rodr{\'{\i}}guez-Ardila et al.\ (2000) demonstrated that
spectra with a high S/N ratio (the exact value is unclear but likely
to be around 30/pixel) and a moderate resolution ($340~km~s^{-1}$
FWHM at $H\alpha$) are sufficient for the decomposition of the
permitted emission-lines in NLS1s. In comparison, the SDSS spectra
used in this work have a somewhat lower S/N on average but better
resolution (estimated to be $132\pm12~km~s^{-1}$ FWHM at $H\alpha$,
cf.\ Greene \& Ho 2005); the effect of the latter compensates the
effect of the former. Spectral decomposition was also carried out in
VVG01, in which the spectral resolution ($FWHM\sim 3.4~\AA$) was
actually a little bit worse than that of the SDSS spectra. We
therefore anticipate that the spectral quality of the SDSS data is
adequate for our purposes, at least for the majority of objects. A
detailed account of spectral-line fitting is presented in
\S\,\ref{sect:linefit}.

{\em Spectral fitting procedure:} In the processing of this large
data set we employed a two-step iteration procedure, which is
outlined below and detailed in the following two subsections. The
continua of the broad emission-line AGN candidates were first
decomposed into starlight and nuclear components. Beforehand, two
kinds of spectral regions had been masked out: firstly, bad pixels
as flagged by the SDSS pipeline; secondly, the wavelength ranges
which may be seriously affected by prominent emission lines
characteristic of QSO spectra. To determine the latter, the
composite SDSS QSO spectrum (Vanden Berk et al.\ 2001) was used
during the first iteration. A pseudo-continuum was fitted and
subtracted, which was chosen as a non-negative linear combination of
properly generated templates taking into account the contribution
from both starlight and nucleus (\S\,\ref{sect:starlight}). The
leftover emission-line spectrum was fitted with Gaussian and/or
Lorentzian profiles, separating broad and narrow line components
(\S\,\ref{sect:linefit}).
%%% above line added by W.Y. (06/02/22)
Then the {\em measured} emission line spectrum was used to replace
the composite SDSS spectrum in the second iteration. This procedure
was reiterated until the fitted parameters of both the continuum and
the emission-line spectrum converged to an acceptable accuracy.

\subsection{Decomposing Starlight and Nuclear Continuum}
\label{sect:starlight}

We modeled the pseudo-continuum of the 387 483 SDSS galaxies and
QSOs with 2 to 4 components
\begin{equation}\label{eq1}
S(\lambda)=A_{host}(E_{B-V}^{host},\lambda)~A(\lambda)+A_{nucleus}(E_{B-V}^{nucleus},\lambda)~[bB(\lambda)+cC(\lambda)+dD(\lambda)]
\end{equation}
where $S(\lambda)$ is the observed spectrum.
$A(\lambda)=\sum_{i=1}^6 a_{i}~IC_{i}(\lambda,\sigma_{*})$
represents the starlight component modeled by our 6 synthesized
galaxy templates, which had been built up from the spectral template
library of Simple Stellar Populations (SSPs) of Bruzual \& Charlot
(2003, hereafter, BC03) using our new method based on the Ensemble
Learning Independent Component Analysis (EL-ICA\footnote{ICA is a
new Blind Source Separation (BSS) method for finding underlying
statistically independent, non-Gaussian  components from
multidimensional data. This statistical and computational technique
can be regarded as an extension to Principle Component Analysis
(PCA) but is much more powerful than the latter. The Ensemble
Learning algorithm is an approach to find an analytic approximation
applicable to the ICA model (Miskin \& MacKay 2001). }) algorithm.
The merits of these galaxy templates are that, firstly, their
physical meanings are clear; secondly, in most cases, an over-fit to
the spectra can be avoided because the most prominent features of
the BC03 SSP library are embodied to the greatest extent in only 6
non-negative independent components (ICs). The details of the galaxy
templates and their applications are presented in Lu et al.\ (2006).
$A(\lambda)$ was broadened by convolving with a Gaussian of width
$\sigma_{*}$  to match the stellar velocity dispersion of the host
galaxy. The un-reddened nuclear continuum is assumed to be
$B(\lambda)=\lambda^{-1.7}$ as given in Francis (1996). $C(\lambda)$
denotes the optical FeII templates of V{\'e}ron-Cetty et al.\ (2004)
with the width fixed to that of the broad component of $H\beta$.
$D(\lambda)$ represents the templates of higher-order Balmer
emission lines ($10\leq n\leq 50$) and Balmer continuum generated in
the same way as in Dietrich et al.\ (2003).
$A_{host}(E_{B-V}^{host},\lambda)$ and
$A_{nucleus}(E_{B-V}^{nucleus},\lambda)$ are the color excesses due
to the extinction of the host galaxy and the dusty torus,
respectively, assuming the extinction curve for the Small Magellanic
Cloud (SMC) of Pei (1992). The fitting was performed by minimizing
the $\chi^{2}$ with $E_{B-V}^{host}$, $E_{B-V}^{nucleus}$, $a_{i}$,
$\sigma_{*}$, $b$, $c$, and $d$ being non-negative free parameters.

To reduce the amount of computation, $E_{B-V}^{host}$,
$E_{B-V}^{nucleus}$, $c$ and $d$ were fixed at zero during the first
iteration. Then the modeled continuum was subtracted from the
observed spectrum and the emission line parameters were measured in
the way described in \S\,\ref{sect:linefit}. $E_{B-V}^{host}$ and
$E_{B-V}^{nucleus}$ were set to be free parameters when the
continuum was fitted the second time. For those objects in which the
$H\beta$ broad emission line is reliably detected (at the $\gtrsim
5~\sigma$ confidence level), $c$ and $d$ were also set free, and the
width of the FeII multiplets was fixed to that of the $H\beta$ broad
component. This time the wavelength ranges of the emission lines to
be masked out were determined using the {\em measured} line
parameters. The iteration continued until the fitted parameters of
both the modeled stellar spectrum and the emission-line spectrum
converged to an acceptable accuracy. Representative examples of the
EL-ICA starlight--nucleus decomposition are shown in
Figure\,\ref{f1}.

\subsection{Emission-Line Fitting}
\label{sect:linefit} The emission-line fitting procedure adopted
here is similar to that used in Dong et al.\ (2005), but with some
modifications in the codes to improve the accuracy of the
measurement of the broad components of H$\beta$ and H$\alpha$, as
described in the following. The emission lines are decomposed into a
broad and a narrow component wherever the latter contributes
non-negligibly. The [NII] doublet are separated from H$\alpha$.

All narrow emission lines, except the [OIII] doublet, were fitted
with a single Gaussian because these lines are often quite weak in
NLS1s  and the S/N is not high for most objects. The [OIII] doublet
were fitted  either with a single Gaussian if the profile is
symmetrical or of low S/N, or with a double Gaussian otherwise. The
flux ratios of the [OIII] and [NII] doublets were fixed at their
theoretical values. The profile and redshift of the H$\alpha$ narrow
component were assumed to be the same as those  of the [NII] and
[SII]  doublets;---this approach is often adopted in separating the
narrow and broad components in the H$\alpha$+[NII]+[SII] regime
(e.g., Ho et al.\ 1997; VVG01; and Dong et al.\ 2005). This
assumption can be tested using type~2 AGNs, where  narrow emission
lines can be  measured much more easily. Here we carried out such a
test using a sample of $\sim 3~000$ type~2 AGNs with high S/N
spectra from SDSS. Figure\,\ref{f2} shows FWHM[NII] (upper left
panel) and FWHM[SII] (upper right panel) against FWHM(H$\alpha$). It
can be seen that both FWHM[NII] and FWHM[SII] are  statistically
well consistent  with FWHM(H$\alpha$). For this type~2 AGN sample,
then, we made a second fit after tying together the profiles and
redshifts of H$\alpha$, [NII]$\lambda\lambda6548,6583$, and
[SII]$\lambda\lambda6716,6731$. The fitted line width  was compared
with that of H$\beta$, and a good agreement was found
(Figure\,\ref{f2} lower left panel). FWHM(H$\beta$) was  also found
to be well correlated with FWHM[OIII], though the scatter is much
larger than in the FWHM(H$\beta$)--FWHM[NII] correlation. We
therefore consider it a good approximation to tie the profile and
redshift of the  narrow component of H$\beta$---in the case of weak
lines, either narrow or broad---to the fitted values of
H$\alpha$+[NII] or [OIII]. We refer to these schemes as ``H$\alpha$
to H$\beta$" and ``[OIII] to H$\beta$", respectively.

In the first place, each of the forbidden and permitted lines was
fitted with a single Gaussian. Objects were assigned as candidates
of broad emission-line AGNs if their Balmer emission line(s) were
significantly broader than the forbidden lines or their FeII
multiplets were detected at the $\gtrsim 5~\sigma$ confidence level.
The spectra of more than 6~000 candidates from the SDSS ``galaxy"
catalog were visually inspected and 1~926 turned out to be broad
line AGNs. These, together with 12~824 objects with $z<0.8$ in the
SDSS ``QSO" catalog, were carried over into the second iteration
stage. After the continuum was subtracted, the H$\alpha$ and
H$\beta$ emission lines were fitted with two Gaussians, one for the
narrow and one for the broad component. As mentioned above, the
width and redshift of the narrow component of H$\alpha$ were tied to
those of the [NII] and [SII] doublets; and for the narrow component
of H$\beta$, the ``H$\alpha$ to H$\beta$" (for $z<0.39$) or ``[OIII]
to H$\beta$" (for $0.39<z<0.8$) schemes were adopted. The objects
were reserved whose broad component(s) of H$\alpha$ or H$\beta$ were
detected at the $>10~\sigma$ level with the FWHM less than
$3~000~km~s^{-1}$. All the rejected objects were double-checked for
confirmation by visual inspection to ensure that no true NLS1s were
missed (actually such checks were made throughout the whole
selection process whenever an object was to be rejected). We also
rejected at this stage the narrow emission line galaxies that were
mis-classified as ``QSOs" by the SDSS pipeline. At the stage of the
third iteration, the broad components of H$\alpha$ and H$\beta$ were
fitted with a Lorentzian profile, while the narrow lines were fitted
in the same way as in the second iteration. Only those having
FWHM(H$\alpha$) or FWHM(H$\beta$) less than 2~500\,km\,s$^{-1}$ were
carried over into the last iteration stage. The emission lines were
then fitted using one of the following four schemes in order of
decreasing preference: 1) the broad and narrow components of
H$\beta$ are de-blended freely; 2) the ``H$\alpha$ to H$\beta$"
scheme; 3) the ``[OIII] to H$\beta$" scheme; and  4) H$\alpha$ and
H$\beta$ lines are fitted with a single Lorentzian (for objects
having no isolated narrow emission line detected or having too noisy
spectra). The fitting results were visually inspected before
acceptance. Only 42 objects could not be fitted satisfactorily by
using the above procedure. Among those exceptions, the majority have
the broad components of H$\alpha$ or H$\beta$ too narrow to separate
from the narrow components; the remaining have their redshifts
incorrectly assigned by the SDSS pipeline. Their spectra were fitted
manually. Figure\,\ref{f3} shows a few examples of emission-line
decomposition in the H$\alpha$ and H$\beta$ wavelength ranges.

\section{Sample Selection}

\subsection{Selection Criteria}

Based on the emission line parameters measured, we compiled our NLS1
sample. The only criterion we adopted for identifying a NLS1 galaxy
is: the ``broad" component of H$\beta$ or H$\alpha$ is detected (at
the $>10~\sigma$ confidence level) and is narrower than
$2~200~km~s^{-1}$ in FWHM. This criterion resulted in 2~011 NLS1
candidates, whose main spectral parameters are summarized in
Table\,\ref{tbl-1}.

It is noticeable that our selection criterion is somewhat different
from the ``conventional'' NLS1 definition (see the Introduction) in
two aspects: firstly, the line width cut-off of $2~200~km~s^{-1}$ is
slightly higher than the commonly used $2~000~km~s^{-1}$; secondly,
we did not restrict ourselves to objects with $[OIII]/H\beta<3$.
These modifications were made based on the following considerations:
(1) The line width distribution for the Balmer lines reveals a
rather smooth transition between NLS1s and BLS1s, which means that
any dividing line is somewhat arbitrary and `operational'. Besides,
different line profiles were used by the different authors (see
VVG01 and the references therein), and sometimes the narrow
component was not taken into account properly (e.g., WPM02),
resulting in underestimation of the line width (see \S3.2). As can
be seen from Figure\,\ref{f4} (upper left panels), typical relative
errors of the broad components of H$\beta$ and H$\alpha$ are 6\% and
3\%, respectively. By relaxing the FWHM cut-off in H$\beta$ or
H$\alpha$ up to $2~200~km~s^{-1}$, we could cover most of the bona
fide NLS1s with low spectral quality, at the expense of introducing
a few spurious ones though. Imposing the stricter criterion of
FWHM=$2~000~km~s^{-1}$ would reduce the sample size to 1~885
objects. (2) The restriction of $[OIII]/H\beta<3$ was originally
introduced to separate NLS1s from Seyfert\,2s where the resolution
and/or S/N of the observed spectra were too low to do so. This is
not the case for at least the majority of objects in our study,
however, as discussed in \S\,\ref{sect:overview}. Consequently, this
criterion becomes no long necessary. In fact, there is only one
object (SDSS\,J104755.93+073951.2, see Figure\,\ref{f5} upper
panels) that fails to meet this restriction.

As a serendipitous finding, one object (SDSS~J113323.97+550415.8)
having the prominent, narrow ``broad" Balmer emission lines
[$FWHM(H\beta)\sim2~100~km~s^{-1}$ and
$FWHM(H\alpha)\sim1~900~km~s^{-1}$] characteristic of NLS1 was
actually a type~2 supernova. Its spectrum and SDSS image are shown
in Figure\,\ref{f5} (lower panel). The object is located at the edge
of the nearby blue compact galaxy UGCA 239
(SDSS~J113323.46+550420.6) and has the same redshift as the galaxy.
It was noticed because its brightness decreased by $\sim 3$
magnitudes within less than one-year between the photometric and
spectroscopic observations. Its disappearance from our later imaging
observation of UGCA 239 using the 2.16m telescope of Beijing
Observatory in January, 2005 confirmed its supernova nature.
Accordingly we removed this object from our NLS1 sample.

\subsection{Reliability of the Sample Parameters and Comparison with Previous Samples}

Our NLS1s were drawn from a well-defined parent population of broad
emission-line AGNs comprising candidates of both QSOs (Richards et
al. 2002) and galaxies (Strauss et al.\ 2002 and Eisenstein et al.\
2001). Considering the fact that the vast majority of the sample
were selected based on the optical selection criteria, we regard the
sample simply as an optically selected sample. The emission-line
parameters were measured with high accuracy by precise subtraction
of the stellar and nuclear continuum and proper decomposition of the
permitted emission-lines into a narrow and a broad component. Since
the broad components of H$\alpha$ and H$\beta$ were measured
separately, we could check our emission-line fitting algorithm by
comparing the two measurements. Figure\,\ref{f4} which plots
FWHM(H$\alpha$) against FWHM(H$\beta$) confirms  the previously
known correlation between the two. Our linear fit, $FWHM(H\alpha) =
(0.842\pm0.016) ~ FWHM(H\beta)$, is consistent with the
$FWHM(H\alpha) = 0.873 ~ FWHM(H\beta)$ found by Eracleous \& Halpern
(1994), and the $1~\sigma$ deviation about $10\%$ is only slightly
larger than typical measuring errors (see the inset panel). We also
compared the fluxes of the broad components of H$\alpha$ and
H$\beta$ and found $F(H\alpha)=(3.028\pm0.017)~F(H\beta)$ and a
dispersion of $\sim 0.36 $ in the H$\alpha$ to H$\beta$ ratio
(Figure\,\ref{f6}). The good agreement between the independent
measurements of the broad H$\alpha$ and H$\beta$ components
indicates that our emission-line fitting algorithm should be
accurate and reliable.

The method for identifying NLS1s we used in this work is largely
different from that used in WPM02; moreover, the WPM02 sample was
drawn merely from the ``QSO" database of the SDSS early data release
(EDR). It would  therefore be interesting to compare our sample with
that of WPM02. We found that the 150 NLS1 candidates in the WPM02
sample are all included in our parent sample of broad emission-line
AGNs. master sample. Our measured FWHM(H$\beta$) are compared with
those in WPM02 in the lower left panel of Figure\,\ref{f4}. For
those objects in which the narrow component of H$\beta$ is
negligible, we found excellent agreement between the two
measurements. However, for about 1/3 of the objects, our
measurements of FWHM(H$\beta$) are significantly larger than theirs.
Going through these objects by visual inspection of their spectra
revealed that the discrepancy is due to the non-negligible
contribution of the H$\beta$ narrow component, which was not taken
into account in WPM02. To demonstrate this effect we plot in
Figure\,\ref{f4} (lower right panel) the FWHM(H$\beta$) ratio of our
measurements to  theirs against the height ratio of the narrow to
broad component of H$\beta$, taken as an indicator of the relative
strength of the narrow component. It can be seen that the measured
FWHM(H$\beta$) values in WPM02 start to deviate from ours when the
narrow component becomes non-negligible, that the deviation gets
worse as the narrow component increases, to reach a factor of 2 or
more when the narrow component surpasses the broad component in
height.

As a result, 5 objects in the WPM02 sample were excluded from our
NLS1 sample because their FWHMs of the H$\beta$ broad component
actually exceeded our criterion of $2~200~km~s^{-1}$. Their
de-blended components of  H$\beta$ and  H$\alpha$ resulting from our
analysis are  shown in Figure\,\ref{f7}. On the other hand, our
sample includes 86 objects from the EDR database that  were not
included in the WPM02 sample. Of these, only 17 fall within the
range defined by the differing cutoffs, $2~000~km~s^{-1}\lesssim
FWHM(H\beta)\lesssim 2~200~km~s^{-1}$, while most of these because
they were  assigned by WPM02 to a separate category, Seyfert 1.5, by
virtue of their  prominent narrow components of the Balmer
emission-lines. We included these simply because their broad
components had FWHM(H$\alpha$) or FWHM(H$\beta$) that satisfied our
cut-off criterion. We argue that the nature of these objects (see
Figure \ref{f8} for some examples) should not be much different from
classic NLS1s, and there seem to be no particular reasons to exclude
them based merely on the prominence of their narrow emission-lines.
In addition, thanks to our starlight--nucleus decomposition and
automated spectral fitting algorithm, our sample includes 13 NLS1s
drawn from the EDR {\em Galaxy} database, which was not considered
in WPM02 at all; in fact, a total of 138 NLS1s were culled from the
Galaxy catalog from the SDSS  DR3 database. The rest few objects not
included in the WPM02 sample have  either low spectral quality or
large uncertainties in the measured spectral parameters. We
accordingly believe that our NLS1 sample has a greater  degree of
completeness than the WPM02 sample.

The following caveats should be noted: Given the quality
$S/N\sim10$ typical of the SDSS spectra for low redshift broad-line
AGNs (BLAGNs), objects with very narrow ``broad" permitted emission
lines could not be effectively identified unless the FeII multiplets
were sufficiently strong. Also, NLS1s with large extinction may have
been missed out. Observations in multi-wavebands, such as the hard
X-ray band, are needed to identify such objects.

\section{Sample Properties}

\subsection{The Fraction of NLS1s}
\label{sect:fraction} Our sample of NLS1s is more than ten times the
size of WPM02, while equally or even more complete and uniform. In
this section we present some basic properties of the NLS1
population, taking advantage of this unprecedented sample size. The
NLS1s as a whole make up 14\% of all the broad emission-line AGNs.
This fraction is consistent with the often quoted value of 15\%,
first suggested by Osterbrock (1988) and  later confirmed by WPM02.
Figure\,\ref{f9} shows the distribution of NLS1s and ``normal"
BLAGNs in redshift and nuclear luminosity (as represented by the
g'-band absolute psf magnitude), as well as the NLS1 fraction within
overall BLAGNs. The NLS1 fraction appears to increase slightly up to
a redshift of 0.45, after which it decreases dramatically toward
high redshift. This is actually not a reflection of the cosmic
evolution of the NLS1 fraction but is rather one of its dependence
on the nuclear luminosity (as seen in Figure\,\ref{f9}), which is
strongly correlated with redshift for a flux-limited sample.

The fast decrease of the NLS1 fraction with increasing nuclear
luminosity in the high luminosity range results actually from the
emission line width--luminosity  relation, which is discussed below
in \S\,\ref{sect:lum-width}. The strong positive correlation of the
NLS1 fraction with luminosity in the low luminosity range
$M_{g}\lesssim -22^{m}$ is unexpected. The fraction  peaks at
$M_{g}\sim -22^{m}$, where NLS1s make up about $20\%$ of all SDSS
BLAGNs; this peak apparently  results from a dip in the luminosity
distribution of BLAGNs here (see Figure\,\ref{f9}), without there
being a corresponding one in that of NLS1s. It is likely that this
dip in the  BLAGNs distribution is due to selection effects, in the
sense that most of low luminosity AGNs were targeted as ``galaxies"
while high luminosity ones as ``QSOs" in the SDSS. However, we argue
that the peak in the NLS1 fraction should be real because objects of
both the NLS1 and BLS1 types were selected according to the same
criteria in the SDSS. It may be a result of the combination of the
local black hole mass function and the distribution of the mass
accretion rate. Detailed discussion on this issue is beyond the
scope of this paper and will be presented elsewhere.

It was once thought that NLS1s are radio quiet; however, the past
few years saw the discovery of about a dozen ``radio-loud" NLS1s
(Siebert et al.\ 1999; Grupe et al.\ 2000; Zhou \& Wang 2002; Zhou
et al.\ 2003). Two objects among these, both identified from the
SDSS, are very-radio-loud (Zhou et al. 2003; Zhou et al. 2005)
according to the criterion given in Sulentic et al. (2003). The
incidence of radio-loud NLS1s, or the dependence of the fraction of
NLS1s on the radio properties is of particular interest. This is
because radio-loud NLS1s show some extreme characteristics opposite
to typical radio-loud AGNs, e.g.\ the extremely steep soft X-ray
spectrum in contrast to the usual flat spectrum (Sulentic et al.\
2000). This issue can be examined in more detail with our large NLS1
sample. Out of the 2 011 NLS1s, 142 were detected in the FIRST (the
Faint Images of the Radio Sky at 20 cm survey, Becker et al. 1995).
The radio detection rate is $7.1\pm0.2\%$, which is significantly
less than what we found for BLS1s at $10.2\pm0.1\%$. It is
remarkable that all the radio sources of the NLS1s detected in the
FIRST were unresolved in their radio images. The distribution of
radio-loudness $R$, (defined as the logarithm of the flux ratio
between 1.4 GHz and the optical g-band) is shown in Figure \ref{f12}
(left panels) respectively for the NLS1s and BLS1s detected in the
FIRST. Compared to the BLS1s, the radio-loudness distribution of the
NLS1s drops much faster beyond $R\sim 1$, indicating a decline of
the NLS1 fraction toward high $R$ values. The NLS1 fraction is $\sim
15.5\%$ in radio-quiet and radio-intermediate AGNs ($R \lesssim 1$),
similar to that from other optically selected samples; it falls down
to $\sim 10\%$ in moderately strong radio AGNs ($1\lesssim R
\lesssim 2$) and further down to $\sim 5\%$ in powerful radio AGNs
($R \gtrsim 2$). In particular, we found only 5 very radio-loud
NLS1s in the whole sample, including the two we reported previously,
SDSS J084957.98+510829.0 and SDSS J094857.32+002225.5. These two
very radio-loud NLS1s show flat spectra and dramatic variability in
the radio band, indicating the presence of relativistic jets
pointing toward the observer. In fact, SDSS J084957.98+510829.0 had
been observed before the SDSS and was classified as a BL Lac object
with a wrong redshift (Zhou et al. 2005 and references therein). For
the other three sources no significant radio variability was found
in time spans of several years between the NVSS (the NRAO VLA Sky
Survey, Condon et al. 1998) and the FIRST. With a spectral index of
$\alpha_{r}\simeq 0.4$, the radio spectrum of the most powerful
NLS1, SDSS J104732.68+472532.1, is somewhat steeper for a
blazar-like source. The dependence of the NLS1 fraction on radio
power (Figure\,\ref{f12}, right panels) is similar to that on the
radio-loudness. A detailed account on the radio properties of NLS1s
will be presented in a forthcoming paper of the series.

Historically, X-ray surveys have been an important tool for the
finding and study of NLS1s. For instance, Puchnarewicz et al.\
(1992) found that about 50\% of their soft X-ray selected AGNs were
NLS1s; a similar result was also obtained by Grupe (2004).
However, we found that the overall NLS1 fraction is only $\sim
19.4\pm0.8 \%$ for our optically selected sample. Plotted in Figure
\ref{f13} (left panel) is the dependence of the NLS1 fraction on the
soft X-ray flux density for the BLAGNs with X-ray counterparts in
the RASS; it reveals a strong dependence---the NLS1 fraction
increases dramatically with the X-ray flux. At the faint end, the
NLS1 fraction is $\sim 15\%$, similar to that found in
optically-selected BLAGN samples. At the bright end, it increases up
to $\sim 40\%$, close to that of the soft X-ray selected AGN samples
reported previously; this is where the NLS1 fractions for optically
selected and soft X-ray selected AGN are reconciled. This is not
surprising, as above this high flux level optically selected and
X-ray selected BLAGN samples are largely overlapping. Interestingly,
we also found a trend of decreasing fraction of NLS1s toward high
X-ray luminosity (Figure \ref{f13}, right panel).

\subsection{The Luminosity--Emission Line Width Relation}
\label{sect:lum-width}

It has been claimed by several authors (e.g., Miller et al. 1992;
VVG01) that the nuclear luminosity is correlated with the width of
the broad Balmer emission-lines for BLAGNs, including NLS1s. We plot
in Figure\,\ref{f10} line luminosity versus line width for H$\alpha$
and H$\beta$, respectively,  for our sample. A weak yet significant
correlation is present between the two parameters, despite the small
range of the line width (the Spearman coefficient $r=0.29$ for
FWHM(H$\beta$) and L(H$\beta$), and $r=0.34$ for FWHM(H$\alpha$) and
L(H$\alpha$), corresponding to chance probabilities much less than
0.1\% in both cases). The correlations appear to result from the
presence of an upper boundary, as indicated by the dashed lines in
Figure \ref{f10}. As will be seen below, the limit is likely  caused
by the existence of upper limits  in the accretion rate in units of
the Eddington rate.

The central black hole mass can be inferred from the size and the
velocity dispersion of the broad emission-line region (BLR),
\begin{equation}\label{eq2}
M_{BH}\propto R_{BLR}~V^2,
\end{equation}
provided the BLR is virialized. We can use the width (FWHM) of Balmer
emission-lines to estimate the velocity dispersion of the BLR.
Supposing all AGNs have on average the same ionization parameters,
the same ionizing spectral energy distribution (SED), the same
BLR densities and column densities,
then we would expect a simple scaling relation between the
BLR size and the monochromatic luminosity at 5100 \AA~ (e.g., Peterson 1993),
\begin{equation}\label{eq3}
R_{BLR}\propto [\lambda L_{\lambda}(5100)]^{0.5}.
\end{equation}
Hence we can express the virial mass of the central black hole
mass as
\begin{equation}\label{eq4}
M_{BH}\propto [\lambda
L_{\lambda}(5100)]^{0.5}~[FWHM(H\beta~or~H\alpha)]^2.
\end{equation}
For our NLS1s, we find that $\lambda L_{\lambda}(5100)$ is tightly
correlated with the luminosity of H$\beta$ broad component,
$L(H\beta)$ (Figure \ref{f11}, left panel). A linear fit yields,
\begin{equation}\label{eq5}
log\{\lambda
L_{\lambda}(5100)\}=(8.57\pm0.28)+(0.840\pm0.007)~log\{L(H\beta)\}.
\end{equation}
Also, assuming that all objects have the same SED, we get
\begin{equation}\label{eq6}
L_{bol}\propto \lambda L_{\lambda}(5100)\propto
[L(H\beta~or~H\alpha)]^{0.840}.
\end{equation}
Therefore, for objects emitting at the Eddington luminosity,
\begin{equation}\label{eq7}
L_{bol}=L_{Edd} \propto M_{BH},
\end{equation}
a given value of $FWHM(H\beta~or~H\alpha)$ corresponds to a value
of $L(H\beta~or~H\alpha)$,
\begin{equation}\label{eq8}
L(H\beta~or~H\alpha) = C~[FWHM(H\beta~or~H\alpha)]^{4.762}.
\end{equation}
The exact value of the constant $C$ depends on the shape of the
ionization continuum, the geometry and physical condition of the
BLR, and the cosmology adopted.
This correspondence is visualized in Figure\,\ref{f10}
by the dashed line with the constant $C$ so chosen that $L_{bol}$ is
a few units of $L_{Edd}$.
It can be seen that most of our NLS1s populate below the line.
This result implies that the radiation of the majority of
AGNs including NLS1s cannot be much higher than
a few units of the Eddington luminosity.
The correlation between the broad line width and the nuclear luminosity
naturally explains the decrease of the NLS1 fraction
toward high luminosities reported in \S\ref{sect:fraction}.

A strong correlation between the
luminosity of continuum and [OIII] emission line was also found
for our NLS1s (see right panel of Figure \ref{f11}), though not as
tight as the $L(H\beta) - \lambda L_{\lambda}(5100)$ correlation.
The best linear fit to the data points yields,
\begin{equation}\label{eq9}
log\{\lambda L_{\lambda}(5100)\}=(7.11\pm0.56)+
(0.885\pm0.0133)~log\{L_{[OIII]}\}.
\end{equation}
This relation is useful for estimating the intrinsic luminosity of
narrow emission-line AGNs based on [OIII]
luminosity (e.g., Zakamska et al. 2003). If the central black hole
mass of these type~2 AGNs can be reliably estimated, one can
evaluate their mass accretion rate and search for type~2 counterparts of
NLS1s in those objects with high Eddington  ratios $L/L_{Edd}$.

\subsection{Broad Emission Lines}

Though not a defining property, NLS1s usually show strong optical
FeII emission lines. Out of the 2 011 NLS1s, the FeII multiplets
were detected  in 1 787 objects at the $>3~\sigma$ level.
For the rest the SDSS spectra were too noisy to yield reliable
measurements and the $3~\sigma$ upper limits were calculated (see
Table\,\ref{tbl-1}). The relative strength of the FeII multiplets is
conventionally expressed as the flux ratio of FeII to H$\beta$:
$R_{4570}\equiv \frac{FeII\lambda
\lambda4434-4684}{H\beta}$, where $FeII\lambda \lambda4434-4684$
denotes the flux of the FeII multiplets integrated over the
wavelength range of $4434-4684~\AA$, and H$\beta$ the flux of the
broad component of H$\beta$ (e.g., VVG01). The distribution of
$R_{4570}$ for the NLS1 sample is shown in Figure \ref{f14}.
The average is $R_{4570}=0.82$ and the dispersion at $1~\sigma$
is 0.37 when only the 1 787 objects with reliable FeII detection
were considered.
This is significantly larger than the typical value of $R_{4570}\sim
0.4$ found in normal AGNs (Bergeron \& Kunth 1984).
The inclusion of the non-detections would not
change the results significantly,
since they only account for $\sim 10\% $ of the whole sample.
About a quarter of all the NLS1s are moderately strong FeII
emitters with $R_{4570}\gtrsim 1$, while this fraction is only $\sim
5\%$ in normal AGNs (Lawrence et al. 1988).
We identified 17 new extreme FeII emitters ($R_{4570}\gtrsim 2$)
from the present NLS1 sample,
which doubled the number of known objects of this kind
(VVG01; Zhou et al.\ 2002; Zheng et al.\ 2002).

The merits of our sample are its large size and uniformity, thanks
to which some weak correlations are expected to be identifiable.
Hence, the sample is suited for an examination of whether the
previously known correlations for broad line AGNs can be extended to
smaller line widths. The E1 space involves primarily the
correlations between three parameters: FWHM(H$\beta$), $R_{4570}$,
and soft X-ray spectral index, $\Gamma$. In Figure \ref{f15}
$R_{4570}$ is plotted against the FWHM of the broad components of
H$\beta$ and H$\alpha$, respectively. We found that the known
anti-correlation between $R_{4570}$ and the FWHM of the broad Balmer
line, though  weak, continues to exist in NLS1s. The Spearman
correlation coefficient is r=-0.23 for $R_{4570}$ vs FWHM(H$\beta$)
and r=-0.31 for $R_{4570}$ vs FWHM(H$\alpha$), corresponding to
chance probabilities of $\ll 0.1\%$ in both cases. The correlation
is slightly stronger between $R_{4570}$ and FWHM(H$\alpha$) than
between $R_{4570}$ and FWHM(H$\beta$); this is because, in general,
the measuring uncertainties are smaller in FWHM(H$\alpha$) than in
H$\beta$. Gaskell (1985) suggested that the large $R_{4570}$ in
NLS1s was due to their Balmer emission lines  being weak rather than
their FeII emission being strong; this was confirmed by VVG01 by
showing a moderately strong correlation between H$\beta$ luminosity
and $R_{4570}$ found in some $200$ broad line AGNs including 64
NLS1s. We did not find such a correlation for NLS1s alone, however.
This result indicates that NLS1s form a distinct subclass, at least
as regarding those showing rather broad low-ionization lines, namely
``population B" (Sulentic et al.\ 2000).

Interestingly, strong correlations were found between the
luminosity of the Balmer emission lines and the equivalent widths
of both the Balmer and FeII emission lines (the ``inverse"  Baldwin
effect) for NLS1s. We do not know exactly the origin of this
strong correlation, which is rather weak---if exists at all---in
BLS1s. Contamination from stellar light can be ruled out as we
believe that starlight has been properly subtracted out. Either a
luminosity-dependent SED, for which the optical to UV spectrum
flattens as luminosity increases, or an increase in the covering
factor of BLR with increasing luminosity may explain such a
correlation.

A strong correlation between the ROSAT photon index $\Gamma$ and
FWHM(H$\beta$) has been well established for broad emission-line
AGNs (Laor et al.\ 1997; Wang, Brinkmann, \& Bergeron 1996; Boller
et al.\ 1996). While BLS1s show a rather small scatter in their
photon indices, NLS1s show a larger dispersion of $\Gamma \sim 1-5 $
(Zhou \& Wang 2002). It is not yet clear whether the
$\Gamma$--FWHM(H$\beta$) correlation could extend to very  small
Balmer line widths. A total of 635 NLS1s have X-ray counterparts in
the RASS catalog; however,  only a fraction have enough photon
counts for an estimation of the photon index. We searched for more
data from the ROSAT Source Catalog of pointed observations (RSC,
Voges et al. 1999) in an attempt to improve the X-ray data quality.
X-ray sources that lie within $30''$ of the optical positions were
regarded as real matches. In all 136 RSC sources were found (all
detected in the RASS) and their RSC data were used in preference to
their RASS data, because of their longer exposures. Following
Schartel et al.\ (1996) and Yuan (1998), we calculated the X-ray
photon indices and X-ray fluxes using the two hardness ratios and
the count-rates given in the ROSAT catalogs for an assumed  Galactic
absorption column density.

Figure\,\ref{f17} plots the ROSAT photon indices $\Gamma$ versus the
FWHM of H$\beta$ and H$\alpha$ for the 310 NLS1s, for which $\Gamma$
could be estimated with reasonable certainty. Of particular
interest, there appears to exist a turn-over in the correlation
trend  at FWHM around $1~000~km~s^{-1}$ for both H$\beta$ and
H$\alpha$. For FWHMs above the turn-over ($FWHM\gtrsim
1~000~km~s^{-1}$) $\Gamma$ is significantly anti-correlated with
both FWHM(H$\alpha$) and FWHM(H$\beta$); applying the Spearman test
gave a correlation coefficient $r=-0.24$ and $r=-0.21$, and a chance
probability of $p=0.1\%$ and $p=0.04\%$, for
$\Gamma$--FWHM(H$\alpha$) and $\Gamma$--FWHM(H$\beta$),
respectively. In the FWHM range below  $1~000~km~s^{-1}$, on the
contrary, the trend is reversed, with $r=0.366$, $p=1.7\%$ and
$r=0.316$, $p=1.2\%$ for the $\Gamma$--FWHM(H$\alpha$) and
$\Gamma$--FWHM(H$\beta$) relations, respectively. Objects with the
steepest spectral slopes (the peak in the $\Gamma$ distribution)
were found to have FWHM values around $\sim 1~000~km~s^{-1}$. For
the 61 NLS1s in the $FWHM$ range of $800-1~200~km~s^{-1}$, the mean
photon index is $<\Gamma>=2.9\pm0.4$, with 43 objects ($\sim 70\%$)
have $\Gamma \gtrsim 2.75$ and only 2 objects have $\Gamma \lesssim
2.0$.

\subsection{Narrow Emission Lines}

For 751 NLS1s in our sample the fluxes of the H$\beta$ narrow
component could be reliably measured at the $>3~\sigma$ level; for
most of the rest meaningful upper limits could be obtained. We plot
in Figure \ref{f18} (upper panel) the distribution of the flux ratio
of [OIII]$\lambda5007$ to the narrow component of H$\beta$,
$[OIII]\lambda5007/H\beta^{n}$. For those objects with only upper
limit on the H$\beta$ component, a lower limit of
$[OIII]\lambda5007/H\beta^{n}$ is set at 10 . Objects having
$[OIII]\lambda5007/H\beta^{n}\lesssim 5$ were found in only $\sim
15\%$ of the NLS1s of our sample; and most of those NLS1s with low
$[OIII]\lambda5007/H\beta^{n}$ ratios are actually  ``composite"
objects in which a large fraction of the emission lines come from
HII regions in the host galaxies. This result is consistent with
that of VVG01 but is contrary to that of Rodr{\'{\i}}guez-Ardila et
al.\ (2000). We plot in Figure \ref{f18} (lower panel) the
diagnostic diagram for the NLR of 168 NLS1s, for which the two sets
of narrow emission lines, $[OIII]/H\beta^{nc}$ and
$[NII]/H\alpha^{nc}$ could be reliably measured. About 10\% of these
NLS1s are scattered into the region occupied by star-forming
galaxies; this fraction is also consistent with that of VVG01 (6/64
in their sample). One important implication is that the majority of
the type~2 counterparts of NLS1s are actually Seyfert~2 galaxies.

Using a sample of $\sim 400$ broad emission-line AGNs selected from
the first data release of the SDSS, Boroson (2005) found that the
low-ionization forbidden lines ([OII], [NII] and [SII] doublets) had
mutually consistent redshifts, while  in about half of the AGNs the
[OIII] doublet was blueshifted  with respect to the redshift defined
by the low-ionization forbidden lines. The average blueshift of
[OIII] was 40 $km~s^{-1}$, and the largest shift could reach
$\gtrsim 400~km~s^{-1}$. This phenomenon also appears in our NLS1
sample (Figure\,\ref{f19}, panels a and b). The peaks of the [OIII]
line in $\sim 60$ NLS1s are shifted by $\gtrsim 400~km~s^{-1}$. The
spectrum of one such object is shown in Figure\,\ref{f20} (upper
panel). These so called ``blue outliers" (Zamanov et al.\ 2002) can
provide a valuable tool in the study of the dynamics of NLR and its
possible connection with other AGN components.

We also found a few NLS1s showing double-peaked profiles of
narrow emission-lines (Figure\,\ref{f20}, lower panel).
Possible explanations  of the origin of such profiles
invoke bipolar outflows, a disk-like emission line region, and binary black
holes with separated NLRs (Zhou et al.\ 2004).

\subsection{Notes on Other Rare Type NLS1s}

One remarkable property of this large sample is its diversity.
Rarities could be readily identified, such as very radio-loud NLS1s,
``blue outliers", and objects with double-peaked narrow emission
lines as mentioned above. Below we present several other rare types
of objects, namely ``UV-deficient'' NLS1s, ``FeII-deficient'' NLS1s,
and NLS1s with broad absorption line troughs (BALs). As illustrative
examples, some of their spectra are shown in Figure\,\ref{f21} and
\ref{f22} and \ref{f23}. Rare type objects are very useful for
examining AGN models.

It is generally believed that each NLS1 harbors a relatively small
black hole accreting at a very high accretion rate. According to
this scenario, NLS1s would have a flat spectral shape in the
optical-to-UV band. While this was  found to be indeed the case for
the majority of the sample, some NLS1s show deficient radiation in
the blue part of their SDSS spectra. Most of these ``UV-deficient''
spectra can be explained as due to heavy extinction caused by a
presumed dusty torus and/or dust lane in the host galaxies. In a few
objects, however, this unusual spectrum cannot be explained by
extinction. For example, the UV dip in the spectrum of SDSS
J142033.7+573900.9 (Figure\,\ref{f21}, upper panel) cannot be
entirely accounted for by extinction because the flux ratio of
$H\gamma/H\beta=0.40\pm0.02$ is close to the theoretical value
expected for a spectrum free from extinction.

Another rare class of NLS1s is concerned with the strength of the
FeII multiplets. While  the FeII emission lines are generally found
to be strong in most NLS1s, they are very weak, or even
undetectable, in some objects with very narrow ``broad" Balmer
emission lines (Figure\,\ref{f22}). The closest type to these
objects is dwarf Seyfert 1 galaxies, such as NGC~4395 (e.g.,
Filippenko \& Sargent 1989). The difference between the two classes
lies in that the FeII deficient NLS1s found here have a relatively
high luminosity with respect to the black hole mass, and hence a
very high accretion rate.

NLS1s and broad absorption line (BAL) QSOs share many common
observational properties (see Brandt \& Gallagher 2000 for a
review). This similarity motivated several authors to investigate
possible connections between the two. For instance, there is the
suggestion that BAL QSOs are misaligned NLS1s (Sulentic et al.\
2000). We have identified 6 NLS1s with possible BAL (3 of the
candidates were serendipitously discovered in the SDSS DR4), whose
SDSS spectra are shown in Figure\,\ref{f23}. The facts are
perplexing. On one hand, BAL QSOs are thought to be otherwise normal
QSOs seen nearly edge-on according to the popular unification
models. On the other hand, there is convincing observational
evidence for a small inclination angle for at least some NLS1s
(e.g., Zhou et al.\ 2003). Despite many observed properties of BAL
QSOs can be explained by orientation-based unification models,
evidence begins to mount that polar outflows do exist, at least in
some BAL QSOs (Zhou et al.\ 2006). It appears that the real
situation is much complex than is assumed by the simple models.

\section{Black Hole--Bulge Relation in NLS1s}
\subsection{$M_{BH}-\sigma_{*}$ Relation}

A growing body of evidence accumulated in the past decade suggests
that the evolution of supermassive black holes (SMBHs) and that of
their host galaxies are tightly linked.  Kormendy \& Richstone
(1995) found that SMBH mass, $M_{BH}$,
is correlated with the mass of
the spheroidal component, $M_{bulge}$, in seven nearby galaxies.
This relationship was later confirmed and quantified by Magorrian
et al.\ (1998) based on a sample of 36 nearby galaxies with Hubble
Space Telescope (HST) photometry and ground-based kinematics
data. Ferrarese \& Merritt (2000) and Gebhardt et al.\ (2000)
demonstrated that $M_{BH}$ is tightly correlated with the stellar
velocity dispersion, $\sigma_{*}$. Nelson et al.\ (2004) measured
the bulge stellar velocity dispersion in 14 Seyfert 1 galaxies
whose SMBH masses were determined using the reverberation mapping
technique, and showed that the Seyfert galaxies followed the same
$M_{BH}-\sigma_{*}$ relation as  non-active galaxies.
The tight correlations for both normal and active galaxies indicate that the
most of the SMBH mass was built up in the past by accretion of the
same gas that also formed the stars in the host galaxy spheroid.
It is of particular interest as to when and how SMBH and the surrounding
spheroid of stars settled down at the right masses. A
straightforward approach to investigate the physical link between
the SMBH growth and the bulge formation is to measure the SMBH
masses and bulge properties in high redshift AGNs, and compare
them with those in low redshift AGNs and quiescent galaxies. This
is a very hard undertaking given the observational difficulties.
One alternative is to study young AGNs in the local universe.
It has been suggested that NLS1s are normal Seyfert galaxies in their
early stage of evolution, when the SMBH is growing fast via
accreting mass at a high rate (e.g., Wang, Brinkmann, \& Bergeron
1996; Boller, Brandt \& Fink 1996; Laor et al.\ 1997, Mathur 2000).
In this sense NLS1s may serve as a surrogate for high redshift AGNs (Wang
\& Lu 2001).

A direct probe of the $M_{BH}-\sigma_{*}$ relation is practically
useful since, in principle, the stellar velocity dispersion of a
bulge in nearby active galaxies can be determined more accurately
than the bulge luminosity or compactness (H{\" a}ring \& Rix
2004). In practice, however, width of the [OIII] emission line
is often used in stead of stellar velocity dispersion due to
observational difficulty of the latter. Using such an
approach several groups of authors explored  the
$M_{BH}-\sigma_{*}$ relation for NLS1s recently,
with  conflicting results:
Mathur et al.\ (2001), Grupe \& Mathur (2004),  Bian \&
Zhao (2004), and Botte et al.\ (2004) found that NLS1s showed a
systematically lower $M_{BH}/M_{bulge}$ ratio than normal broad
line AGNs, while Wang \& Lu (2001), Wandel (2002; 2004) did not
find a clear difference between the two.
The difference may be attributed, at least partly,
to the different line widths used by the different authors.
Wang \& Lu (2001) argued that only symmetric line
component can trace the gravitational potential of
the bulge and so they used the width of the blue-wing subtracted [OIII] line
from VVG01; in contrast,  most others adopted the total [OIII]
line profile. It has been shown that the [OIII] line profiles in
most Seyfert galaxies are evidently asymmetric,
so signifying complex
kinetics of the emission line regions (VVG01 and references
therein). Moreover, contamination from the FeII emission
aggravates the problem, which  is especially true for NLS1s. To
remove all these uncertainties from the study of the SMBH-bulge
relation, it is essential to have direct measurements of
$\sigma_{*}$ in stead of the [OIII] line width.
The underlying physical justification is clear, since
the stellar kinematics is more indicative of the
bulge gravitational potential  than is the gas in the NLR of the AGN.
As an attempt along this line, Botte et al.\
(2005) obtained the near-infrared spectra of 8 NLS1s with a
moderately high resolution of $\sim 50~km~s^{-1}$ per pixel, and
measured directly the stellar velocity dispersions using the
CaII$\lambda \lambda$ 8498,8542,8662 absorption triplet. They
found that $\sigma_{[OIII]}\gtrsim \sigma_{*}$ for 8 of the 10
NLS1s with $\sigma_{*}$  available so far, and demonstrated that
the NLS1s follow the same $M_{BH}-\sigma_{*}$ relation as normal broad
line AGNs and non-active galaxies.
However, the sample of Botte et al.\ (2005)
is too small to give conclusive results, as was warned by
the authors themselves.
A large NLS1 sample with properly
measured $\sigma_{*}$ is utterly needed.

As discussed above in \S\,\ref{sect:starlight}, in the
process of decomposing the stellar and nuclear components in NLS1s,
$\sigma_{*}$ can be obtained automatically\footnote{Because the
spectral resolution of the galaxy templates we used is similar to
that of the SDSS, deconvolution of the SDSS spectra was not
necessary to obtain the stellar velocity dispersions.}.
Though the resolution and S/N ratio of the SDSS spectra
are not perfect for exploring the $M_{BH}-\sigma_{*}$ relation in NLS1s,
the present sample is so large as to render it sensitive for identifying
any weak trend. Apart from the spectral resolution and
signal-to-noise ratio, the measurement uncertainty of $\sigma_{*}$
is also strongly dependent on the relative contribution of the
stellar light. We found that $\sigma_{*}$ could be measured with
acceptable errors for spectra with a median signal-to-noise ratio
$S/N\gtrsim 10$ and a relative contribution of starlight more than
40\% of the total flux at 5100\AA. A small number of objects have
measured $\sigma_{*}$ values as small as $\sigma_{*}<50~km~s^{-1}$,
which is actually beyond the resolution of the SDSS; they
were discarded from the analysis.
Adopting these criteria yielded 308 objects with,
we believe, reasonably well determined $\sigma_{*}$;
they formed the subsample which we used to test the black
hole--bulge relation for NLS1s.
Figure\,\ref{f24} plots, for this subsample, the estimated
$M_{BH}$ versus $\sigma_{*}$, with the $M_{BH}$ determined using
the line width--luminosity mass scaling relation (e.g., Kaspi et al.
2000; McLure \& Jarvis 2002; Dietrich \& Hamann 2004),
\begin{equation}\label{eq10}
\frac{M_{BH}}{M_{\odot}}=4.82~10^6~\{\frac{\lambda
L_{\lambda}(5100)}{10^{44}~erg~s^{-1}}\}^{0.7}~\{
\frac{FWHM(H\beta)}{1~000~km~s^{-1}}\}^2.
\end{equation}
The well known $M_{BH}$--$\sigma_{*}$ relation for normal galaxies,
as parameterized by Tremaine et
al.\ (2002),
\begin{equation}\label{eq11}
log(\frac{M_{BH}}{M_{\odot}})=(8.13\pm0.06)+(4.02\pm0.32)~log(\frac{\sigma_{*}}{200~km~s^{-1}}).
\end{equation}
is marked in the plot by the dashed line. A correlation is
clearly present, which follows the trend of the
$M_{BH}$--$\sigma_{*}$ relation for galaxies (the Spearman
correlation coefficient r=0.27 giving a chance probability of $p=2~10^{-6}$).
Meanwhile, the data points of NLS1s show  a systematic
downward shift relative to the relation for galaxies as defined in
Equation \ref{eq5}, with the bulk of the subsample ($\sim 90\%$)
located  below the dashed line.
The results indicate that, firstly,
the correlation between $M_{BH}$ and $\sigma_{*}$ remains valid for NLS1s;
and secondly, the SMBHs in NLS1s are underage and their growth lags
behind the formation of the bulges, as suggested by e.g., Mathur et al.\ (2001).
We stress that our results are actually consistent with
Botte et al.\ (2005);  their much smaller sample
prevented the authors from drawing the same conclusion.
In fact, only 2 of the 10 NLS1s in their sample are located above the
$M_{BH}$--$\sigma_{*}$ line for normal galaxies and normal AGNs.

Caution should be exercised when accepting the above results,
considering that the derived $\sigma_{*}$ may be affected
by contamination by the rotation of the galactic disk.
This issue presents  a matter of concern here since,
for NLS1s at typical redshifts in our sample,
the fiber aperture  would cover the whole galactic disk.
For a further examination of the reality of the
$M_{BH}-\sigma_{*}$ relation obtained above, we
tried to minimize the contamination effect
of the galactic disk rotation.
We visually inspected the SDSS images of NLS1s at
redshifts $<0.1$ with directly measured $\sigma_{*}$, and
picked out objects for which either the host galaxies appear to be face-on
or the $3^{''}$ fiber aperture is dominated by the galactic bulge contribution.
This selection yielded 33 objects; their SDSS images are
shown in Figure\,\ref{f25}.
For these objects, we believe that
the contamination of the disk, even if significant, should not lead
to overestimation of $\sigma_{*}$. Their $M_{BH}-\sigma_{*}$
relation is plotted in Figure\,\ref{f26}. It can be seen that the
same trend of correlation  persists, and all the data points are
distributed well below the expected $M_{BH}-\sigma_{*}$  relation
for galaxies of Tremaine et al.\ (2002).

\subsection{$\sigma_{*}$--Narrow Line Width Relation}

Since we have relatively good measurements of  $\sigma_{*}$ for the
subsample, we could use them as a standard calibrator to test
potential correlations with the width of any narrow emission-lines.
This has practical potential in the sense that in many NLS1s,
especially luminous ones, the measurement of $\sigma_{*}$ is not
available, that maybe the line width could serve as a good
surrogate. Below we test the previously reported relations between
$\sigma_{*}$ and the FWHM of the [NII] and [OIII] lines. We plot in
Figure\,\ref{f27} the FWHM[NII] values (deconvolved to remove the
effect of the SDSS spectral resolution) versus $\sigma_{*}$ for 240
out of the 308 NLS1s, for which the [NII] doublet is detected at the
$>10~\sigma$ level. A significant correlation was found at a
probability level $p=1.8~10^{-11}$ (the Pearson correlation test).
The objects were found to be distributed almost symmetrically with
respect to the often quoted relation $FWHM[NII]\sim 2.35~\sigma_{*}$
with a $1\sigma $ deviation of $\sim 30\%$ (solid line in
Figure\,\ref{f27}). The result indicates that the [NII] emission
line can be used as a surrogate for stellar velocity dispersion in
NLS1s. We extended the above test to a larger sample of Seyfert~2
galaxies compiled from the SDSS, comprising some 3 000 objects with
high spectral S/N ratios. A correlation is found very similar to
that for the NLS1s (Figure\,\ref{f28}). A linear fit yielded
$FWHM[NII]\sim 2.62~\sigma_{*}$, as compared to the often quoted
$FWHM[NII]\sim 2.35~\sigma_{*}$. We conclude that, for Seyfert~2s,
$FWHM[NII]$ can also be used as a surrogate of $\sigma_{*}$. We find
that, after taking all the  scatter into account, $FWHM[NII]$ can
predict black hole masses to within a factor of 3.

In comparison, the [OIII]--$\sigma_{*}$ correlation is only
marginal. Our result suggests that the [OIII] line width, which has
been widely used in the literature, is actually not a good indicator
of the $\sigma_{*}$ and,  hence, of the black hole mass. In fact,
this is not unexpected considering that, in more than half of the
NLS1s in our sample, the [OIII] lines have complex profiles. In a
systematic investigation of the stellar and gaseous kinematics in
narrow emission-line AGNs, Greene \& Ho (2005) also noted the
different behavior of the low-ionization lines and [OIII], and
pointed out that $\sigma_{[OIII]}$ cannot be  directly used to
replace $\sigma_{*}$. Using a two-component Gaussian fit to the
[OIII] line profiles, the authors found that the core of the [OIII]
line could trace $\sigma_{*}$ statistically  but with a considerable
scatter. This approach, however, cannot  yield meaningful results
for most of the objects in our sample because of the low S/N ratio
and/or the effect of the residuals of the FeII emission in their
SDSS spectra.

Having established the good $FWHM[NII]$--$\sigma_{*}$ relation, we
can extend the above examination to the black hole--bulge relation
by making use of FWHM[NII]. The advantage of so doing is that,
compared to $\sigma_{*}$, FWHM[NII] is much easier to measure and is
available for a larger number of NLS1s. In Figure\,\ref{f29}, we
plot $M_{BH}$ versus FWHM[NII] for 613 NLS1s with  reliably measured
FWHM[NII]. A correlation is present, but there is a systematic
deviation from the equivalent relation expected for galaxies given
by Tremaine et al.\ (2002) transformed from
$FWHM[NII]=2.35~\sigma_{*}$ (dashed line). The results confirm those
obtained above for a smaller sample by making direct use of
$\sigma_{*}$ (Figure\,\ref{f24}).

\section{Conclusions and Prospects}
\subsection{Summary of Conclusions}

We have systematically analyzed the spectroscopic data from the QSO
and galaxy database in the SDSS DR3. After proper subtraction of the
stellar and nuclear continua as well as the FeII multiplets,
prominent emission lines were carefully modeled and measured with
typical relative errors less than 10\%. A large sample of $\sim
2~000$ NLS1s has been selected based on a simple well-defined
criterion: the presence of a ``broad" component of the H$\beta$ (or
H$\alpha$) line with $FWHM\lesssim 2~000~km~s^{-1}$. The present
sample outnumbers previous NLS1 samples by a factor of $\sim 10$.
Taking advantage of such an unprecedented large and uniform sample
with accurately measured spectral parameters, we carried out various
statistical analyses, some of which were possible only for the first
time.

Overall, the often quoted frequency $\sim 15\%$ is confirmed for
finding NLS1s in optically selected broad emission-line AGNs;
meanwhile, the frequency is also found to be strongly dependent on
the optical luminosity. The fraction of NLS1s peaks around
$M_{g}\sim-22^{m}$ (the SDSS g'-band absolute magnitude) and drops
quickly toward both the high- and low-luminosity ends. We interpret
the lower chance of finding NLS1s in high luminosity AGNs as a
result of an imposed upper limit on the Eddington ratio
($L/L_{Edd}$), of a few units. The smaller fraction in the low
luminosity range may arise from the combination of two factors: the
shape of the SMBH mass function and the distribution of the
accretion rate. Low luminosity NLS1s contain a very small SMBH mass
accreting at a very high rate, which are rare in the local universe.
We do not confirm the previous results that $\sim 50\%$ of soft
X-ray (e.g. ROSAT) selected AGNs are NLS1s. Such a fraction is found
only for X-ray bright AGNs, but drops quickly with decreasing soft
X-ray flux. The frequency of finding NLS1s in faint ROSAT AGNs
($\lesssim 10^{-12.5}~erg~s^{-1}~cm^{-2}$ in the 0.1--2.4\,keV band)
is actually the same as in optically selected AGNs. The dependence
of the NLS1 fraction on the soft X-ray luminosity shows the same
trend as on the optical luminosity, and the fraction peaks around
$\sim 10^{43.2}~erg~s^{-1}$. Deeper X-ray surveys than the RASS are
needed to pin down the exact peak luminosity. Moreover, we find that
the NLS1 fraction is also strongly dependent on the radio-loudness
and the radio luminosity. The frequency of finding NLS1s in
radio-quite AGNs (with logarithmic radio-to-optical flux ratio
$R<1$) is $\sim 15\%$, similar to that for optically selected AGNs.
This value drops to $\sim 10\%$ in moderately strong radio AGNs
($1<R<2$), and further down to $\sim 5\%$ in powerful radio AGNs
($R>2$).

We find that on average the relative strength of the FeII emission
($R_{4570}$) in NLS1s is about twice of that in normal AGNs;
$R_{4570}$ is anti-correlated with the width of the broad component
of the Balmer emission lines. The  well known correlation between
the width of the broad low-ionization line and the soft X-ray
spectral slope is found to hold true for NLS1s in general; in
addition, it shows some quite peculiar behavior: $\Gamma$ is
anti-correlated with FWHM(H$\beta$) in the range
$FWHM(H\beta)\gtrsim 1~000~km~s^{-1}$, then the trend is reversed in
the range $FWHM(H\beta)\lesssim 1~000~km~s^{-1}$. We find that the
equivalent width of the H$\beta$ and FeII emission lines are
strongly correlated with the H$\beta$ and the continuum luminosity.
We do not find any difference in the NLR properties between NLS1s
and normal AGNs. About 10--20\% of NLS1s show abnormally low line
ratios of [OIII] to the narrow component of H$\beta$, which is a
result of the contribution of HII regions in the host galaxies.
Blueshifts of the [OIII] emission line were found in about half of
the NLS1s, similar to that in normal AGNs.

Stellar velocity dispersions $\sigma_{*}$ could be measured within
statistically meaningful uncertainties for 308 NLS1s. For this
subsample we examined the $M_{BH}$--$\sigma_{*}$ relation for NLS1s.
We found that the black hole mass is also significantly correlated
with the stellar velocity dispersion; however, the bulk of the NLS1s
have black hole masses statistically smaller than expected from the
$M_{BH}-\sigma_{*}$ relation for normal galaxies and AGNs. This
results support the hypothesis that NLS1s are underage AGNs, and the
growth of their SMBH lags behind the formation of the galactic
bulge. With a typical relative error $\sim 30\%$, the line width of
[NII] is found to be a better indicator of $\sigma_{*}$ than that of
[OIII]. By using a larger subsample with reliably measured
FWHM[NII], the $M_{BH}$--bulge relation for NLS1s was further
examined and similar results were obtained.

\subsection{Remarks on Future Work}

Limited by the scope of the this paper, most of the analyses
presented above are only phenomenological. Systematic and detailed
investigations based on this large data set of some of the
interesting aspects as outlined in the Introduction (\S\,1) are just
underway. For instance, we have initiated a program to study the
X-ray spectral and temporal properties of the NLS1s in our sample by
making use of archival serendipitous data from ROSAT, XMM-Newton and
Chandra. The multiwavelength SED can be constructed and studied by
compiling data from the existing archives in various wavebands. The
present results on the $M_{BH}-\sigma_{*}$ relation for NLS1s are
only statistically meaningful;  observations with higher spectral
resolution and S/N ratios are needed for more precise results.
Moreover, mounting evidence accumulated in the past decade indicates
that the AGN phenomena and the star formation history are closely
linked; the preliminary result on the $M_{BH}-\sigma_{*}$ relation
for NLS1s may represent an important link in the sequence of galaxy
evolution. In this direction, we are going to study the stellar
content of the host galaxies in the hope of finding any clues to the
starburst--AGN connection.

Some peculiar objects of various rare types have been identified in
our NLS1 sample, namely, very-radio-loud NLS1s, [OIII] ``blue
outliers", double-peaked narrow line emitters, UV-deficient NLS1s,
and BAL NLS1s. Detailed studies of their properties, which are
expected to be useful for constraining AGN models, will be addressed
in forthcoming papers.

In the above and other related studies of NLS1, a major source of
uncertainty comes from the determination of the black hole masses.
Mass estimation using a one-epoch spectrum is subject to large
uncertainties (cf., Vestergaard 2004); reliable measurements of the
SMBH mass are limited to only a few NLS1s so far, however. With
interest we noted that a few percent objects in our NLS1 sample have
duplicated observations in the SDSS, some of which show large
amplitude variability (see Figure\,\ref{f30}). This property
presents us a great potential of measuring the black hole masses of
NLS1s by making use of the much more accurate reverberation mapping
technique. We plan to monitor a NLS1 sample with a moderate size to
select candidates and then to carry out the reverberation mapping.
The work will also enable us to explore the dynamics of the BLR for
NLS1.

\acknowledgments We thank Tao Kiang for reading the manuscript and
correcting the English writing. This work was supported by Chinese
NSF through NSF-10533050 and NSF-10473013, the Bairen Project of
CAS, and a key program of Chinese Science and Technology Ministry.
This paper has made use of the data from the SDSS. Funding for the
creation and the distribution of the SDSS Archive has been provided
by the Alfred P. Sloan Foundation, the Participating Institutions,
the National Aeronautics and Space Administration, the National
Science Foundation, the U.S. Department of Energy, the Japanese
Monbukagakusho, and the Max Planck Society. The SDSS is managed by
the Astrophysical Research Consortium (ARC) for the Participating
Institutions. The Participating Institutions are The University of
Chicago, Fermilab, the Institute for Advanced Study, the Japan
Participation Group, The Johns Hopkins University, Los Alamos
National Laboratory, the Max-Planck-Institute for Astronomy (MPIA),
the Max-Planck-Institute for Astrophysics (MPA), New Mexico State
University, Princeton University, the United States Naval
Observatory, and the University of Washington.

%%%%%%%%
%\end{document}
%%%%%%%% cut off before figures

%%%%%%%%%%%%%%%%%%%%%%%%%%%%%%%%%%%%%%%%%%%%%%%%%%%%%%%%%%%%%%%%%%%
\clearpage \thispagestyle{empty} \tablenum{1} \topmargin 0.0cm
\begin{deluxetable}{rrrrrrrrrrrrrr}
\rotate \tabletypesize{\tiny} %\tabletypesize{\scriptsize}
\tablecaption{Emission-Line Properties of SDSS NLS1s (a portion for
guidance of its form and content). Col. 1: object name in J2000,
col. 2: redshift given by the SDSS spectroscopic pipeline, col. 3:
monochromatic luminosity at 5100 \AA, col. 4: featureless nuclear
continuum fraction at 5100 \AA, col. 5: H$\beta$ narrow component
flux, col. 6: H$\beta$ broad component flux, col. 7: $H\beta$ broad
component FWHM, col. 8: [OIII]$\lambda$5007, col. 9: H$\alpha$
narrow component flux, col. 10: H$\alpha$ broad component flux, col.
11: $H\alpha$ broad component FWHM, col. 12: [NII]$\lambda$6583
flux, col. 13: [NII]$\lambda$6583 FWHM, col. 14: optical FeII
strength relative to $H\beta$ broad component. \label{tbl-1}}
\tablewidth{670.0pt} \tablehead{ \colhead{Object} &
\colhead{Redshift} & \colhead{$\lambda
L_{\lambda,5100}$\tablenotemark{a}} & \colhead{FC\tablenotemark{b}}
& \colhead{F(H$\beta^{nc}$)\tablenotemark{c}} &
\colhead{F(H$\beta^{bc}$)\tablenotemark{c}} &
\colhead{FWHM(H$\beta^{bc}$)\tablenotemark{d}} &
\colhead{F[OIII]\tablenotemark{c}} &
\colhead{F(H$\alpha^{nc}$)\tablenotemark{c}} &
\colhead{F(H$\alpha^{bc}$)\tablenotemark{c}} &
\colhead{FWHM(H$\alpha^{bc}$)\tablenotemark{d}} &
\colhead{F[NII]\tablenotemark{c}} &  \colhead{FWHM[NII]\tablenotemark{d}} &  \colhead{$R_{4570}$} \\
%\colhead{hhmmss.ss ddmmss.s}& \colhead{} &\colhead{$erg~s^{-1}$ }
%& \colhead{ } & \colhead{$10^{-17}~erg~s^{-1}~cm^{-2}$ } &
%\colhead{$10^{-17}~erg~s^{-1}~cm^{-2}$} & \colhead{$km~s^{-1}$} &
%\colhead{$10^{-17}~erg~s^{-1}~cm^{-2}$} &
%\colhead{$10^{-17}~erg~s^{-1}~cm^{-2}$} &
%\colhead{$10^{-17}~erg~s^{-1}~cm^{-2}$} & \colhead{$km~s^{-1}$}
%\colhead{$10^{-17}~erg~s^{-1}~cm^{-2}$} & \colhead{$km~s^{-1}$} & \colhead{}\\
\colhead{(1)} & \colhead{(2)} & \colhead{(3)} & \colhead{(4)} &
\colhead{(5)} & \colhead{(6)} & \colhead{(7)} & \colhead{(8)} &
\colhead{(9)}& \colhead{(10)} & \colhead{(11)}& \colhead{(12)}&
\colhead{(13)} & \colhead{(14)}} \startdata

000011.41$+$145545.7&0.459621&44.28&0.65&2$\pm$1&257$\pm$7&1088$\pm$40&46$\pm$3&&&&&&0.64$\pm$0.06 \\
000109.13$-$004121.7&0.416622&44.33&0.66&28$\pm$6&287$\pm$11&1629$\pm$135&91$\pm$2&&&&&&0.69$\pm$0.07 \\
000154.27$+$000732.5&0.139595&43.31&0.30&97$\pm$4&420$\pm$16&1872$\pm$106&655$\pm$6&386$\pm$7&1269$\pm$15&1614$\pm$31&142$\pm$3&183$\pm$4&0.22$\pm$0.05 \\
000208.83$-$001742.7&0.651891&44.69&0.82&&298$\pm$13&1470$\pm$77&60$\pm$4&&&&&&1.36$\pm$0.12 \\
000257.37$-$090015.0&0.516586&44.42&0.70&&230$\pm$8&959$\pm$44&64$\pm$6&&&&&&1.06$\pm$0.09 \\
000410.81$-$104527.1&0.239706&44.28&0.77&65$\pm$6&697$\pm$14&1137$\pm$39&186$\pm$6&218$\pm$13&2204$\pm$24&1035$\pm$20&147$\pm$5&199$\pm$8&1.49$\pm$0.07 \\
000834.72$+$003156.2&0.263034&44.45&0.78&56$\pm$10&1381$\pm$21&1537$\pm$44&94$\pm$5&160$\pm$24&3467$\pm$38&1302$\pm$25&41$\pm$7&289$\pm$31&1.28$\pm$0.04 \\
000913.79$-$101246.7&0.614361&44.64&0.68&1$\pm$3&257$\pm$14&1380$\pm$110&53$\pm$5&&&&&&1.18$\pm$0.17 \\
001010.03$+$005126.6&0.387000&44.17&0.57&5$\pm$2&143$\pm$10&2014$\pm$183&43$\pm$3&30$\pm$6&400$\pm$26&1779$\pm$164&8$\pm$3&163$\pm$43&1.15$\pm$0.18 \\
001100.32$+$134812.2&0.686029&44.78&0.88&2$\pm$1&384$\pm$13&1468$\pm$68&15$\pm$3&&&&&&0.99$\pm$0.10 \\
001104.85$-$092357.9&0.695986&44.88&0.80&&394$\pm$14&1401$\pm$64&26$\pm$5&&&&&&1.09$\pm$0.10 \\
001137.25$+$144201.4&0.131834&43.73&0.50&183$\pm$7&1011$\pm$24&2082$\pm$78&399$\pm$7&926$\pm$18&3621$\pm$35&1796$\pm$27&594$\pm$10&274$\pm$4&0.40$\pm$0.03 \\
001416.92$+$145038.4&0.206182&43.91&0.74&11$\pm$8&749$\pm$16&1465$\pm$66&228$\pm$5&32$\pm$18&2050$\pm$26&1116$\pm$29&90$\pm$7&358$\pm$25&0.97$\pm$0.05 \\
001441.54$+$154400.5&0.358144&44.12&0.41&5$\pm$3&228$\pm$10&1707$\pm$116&46$\pm$2&37$\pm$9&773$\pm$33&1755$\pm$114&31$\pm$6&244$\pm$48&0.86$\pm$0.12 \\
001457.47$-$094511.0&0.177650&43.23&0.32&16$\pm$5&221$\pm$12&1410$\pm$146&164$\pm$4&108$\pm$10&964$\pm$15&1361$\pm$43&22$\pm$3&294$\pm$20&0.10$\pm$0.08 \\

\enddata

\tablenotetext{a}{~In units of $erg~s^{-1}$.}
\tablenotetext{a}{~The nuclear light fraction is unreliable for
$FC\lesssim0.25$.} \tablenotetext{c}{~Observed frame line flux in
units of $10^{-17}~erg~s^{-1}~cm^{-2}$.} \tablenotetext{d}{~In
units of $km~s^{-1}$.}
\end{deluxetable}

\clearpage

\setcounter{figure}{0}
\begin{figure*}[h!]
\centering 
\caption{SEE ATTACHED JPG FIGURE Representative examples
of the EL-ICA starlight-nucleus decomposition. In each
panel, we plot the observed spectrum (black), the combination of
the models (red), the decomposed components of the host galaxy
(blue) and the power law continuum of the nucleus (green), and the
FeII multiplets (pink). The observed spectra are
smoothed with a boxcar of 5 pixels for the illustration.} \label{f1}
\end{figure*}

\setcounter{figure}{1}
\begin{figure*}[h!]
\centering  \caption{SEE ATTACHED JPG FIGURE Comparison of the narrow
emission-line widths for $\sim 3~000$ high S/N type 2 AGNs. In the
upper two panels, H$\alpha $, [NII] and [SII] lines are all fitted
with a single Gaussian. The profiles and redshifts of the [NII] doublet
are tied, and so are the [SII] doublet. The flux ratio of the
[NII] doublet is fixed to its theoretical value. It can be seen
that FWHM($H\alpha $) agree well with FWHM[NII] and FWHM[SII]. In
the lower left panel, FWHM[NII] is compared to FWHM(H$\beta$).
The profiles and redshifts of the [NII] doublet are tied to that of
H$\alpha$ and [SII] during the measurement of these lines.
It can be seen that FWHM(H$\beta$) agrees with FWHM[NII] within their
respective uncertainties. FWHM(H$\beta$) is plotted against FWHM[OIII] in the
lower right panel. The scatter is much larger than those in the other three
relations. } \label{f2}
\end{figure*}

\setcounter{figure}{2}
\begin{figure*}[h!]
\centering   
\caption{SEE ATTACHED JPG FIGURE 
Sample results of the
de-blending procedure applied to the H$\beta$+[OIII] (left panels)
and H$\alpha$+[NII] (right panels) ranges. } \label{f3}
\end{figure*}

\setcounter{figure}{3}
\begin{figure*}[h!]
\centering  \caption{SEE ATTACHED JPG FIGURE Distribution of the
relative measurement uncertainty of the H$\beta$ and H$\alpha$ broad
component FWHM (upper left panel histograms). The correlation
between the FWHM(H$\alpha$) and FWHM(H$\beta$) of 1 220 NLS1s with
$z\lesssim 0.4$ in our sample are plotted in the upper right
panel. The dash line denotes the best linear fit,
$FWHM(H\alpha)=0.861 FWHM(H\beta)$.
The distribution of the discrepancy between the measured and
the expected values is plotted as the inset.
The scatter of $\sigma \sim 10\%$
is only a little larger than the measurement uncertainty of
FWHM(H$\alpha$) and FWHM(H$\beta$) (5.5\% and 2.6\%, respectively).
Our measurements of FWHM(H$\beta$) are compared to those of WPM02
in the lower left panel.
In the lower right panel, we plot the
FWHM(H$\beta$) ratio of our measurement to that of WPM02 against
the height ratio of H$\beta$ narrow to broad component. Objects
with detected H$\beta$ narrow components are denoted by filled
circles.
Unfilled circles denote those in which narrow
component of H$\beta$ is not detected. Note the concordance
between our measurements and those in WPM02 for those objects in
which the narrow component of H$\beta$ emission-line is
negligible. The difference between our measurements and those in
WPM02 in a handful of objects is due to the contribution of the
H$\beta$ narrow component, which was not considered by the
WPM02 authors.} \label{f4}
\end{figure*}

\setcounter{figure}{4}
\begin{figure*}[h!]
 \centering  \caption{SEE ATTACHED JPG FIGURE The upper panel shows the
spectrum of SDSS J104755.93+073951.2, the only object in our NLS1
sample with $[OIII]/H\beta>3$ and broad component of H$\alpha$ and
H$\beta$ less than $2~000~km~s^{-1}$. The broad permitted emission
lines are illustrated on an expanded scale in the
middle panel with the error array plotted in thin curve. The lower panel shows
the newly identified type 2 SNe that fulfils our criterion of
NLS1s. It SDSS image is displayed in the inset panel. }
\label{f5}
\end{figure*}

\setcounter{figure}{5}
\begin{figure*}[h!]
 \centering  \caption{SEE ATTACHED JPG FIGURE Correlation between
fluxes of the H$\alpha$ and H$\beta$ broad components. The dash line
denotes the best linear fit to the data. Distribution of the Balmer
decrement, the flux ratio of the H$\alpha$ to H$\beta$ broad component,
is shown in the
inset. } \label{f6}
\end{figure*}

\clearpage
\setcounter{figure}{6}
\begin{figure*}[h!]
\caption{SEE ATTACHED JPG FIGURE Same as Figure \ref{f2}
but for the 5 objects which were classified as NLS1s in WPM02
but rejected as such in this paper. } \label{f7}
\end{figure*}

\setcounter{figure}{7}
\begin{figure*}[h!]
 \centering  \caption{SEE ATTACHED JPG FIGURE Same as Figure \ref{f2}
but for the 14 of the 86 objects classified in this paper as NLS1
candidates but not included in the sample of WPM02.
} \label{f8}
\end{figure*}

\setcounter{figure}{8}
\begin{figure*}[h!]
 \centering  \caption{SEE ATTACHED JPG FIGURE Left panels: redshift and
luminosity distribution of the NLS1 sample (hatched area) and the
parent BLAGN sample (open area). Right panels: dependence of the
NLS1 fraction on redshift and luminosity. } \label{f9}
\end{figure*}

\setcounter{figure}{9}
\begin{figure*}[h!]
 \centering  \caption{SEE ATTACHED JPG FIGURE Luminosity
versus FWHM for the broad components of the H$\beta $ and H$\alpha$
emission lines, respectively. Note that the vast majority of NLS1s populate
in the region below the guiding lines (dashed)
which represent a constant luminosity
a few units of the Eddington luminosity.
See \S\,\ref{sect:lum-width} for detailed discussion. }
\label{f10}
\end{figure*}

\setcounter{figure}{10}
\begin{figure*}[h!]
 \centering  \caption{SEE ATTACHED JPG FIGURE Correlation between
the nuclear monochromatic luminosity at 5100 \AA
with  the H$\beta$ (left panel) luminosity and [OIII] (right panel)
luminosity. The solid line
is the best linear fit. } \label{f11}
\end{figure*}

\setcounter{figure}{11}
\begin{figure*}[h!]
 \centering  \caption{SEE ATTACHED JPG FIGURE Distribution of
the radio-loudness (left panels) and radio power (right panels) of
NLS1 and BLS1s. } \label{f12}
\end{figure*}

\setcounter{figure}{12}
\begin{figure*}[h!]
 \centering  \caption{SEE ATTACHED JPG FIGURE Dependence of the
fraction of NLS1s on the X-ray flux (left panels) and X-ray luminosity
(right panels). } \label{f13}
\end{figure*}

\setcounter{figure}{13}
\begin{figure*}[h!]
 \centering  \caption{SEE ATTACHED JPG FIGURE Distribution of the
relative strength of the FeII multiplets, $R_{4570}$, for the NLS1
sample. Objects with $3~\sigma$ upper limits are
shown by the hatched histogram.} \label{f14}
\end{figure*}

\setcounter{figure}{14}
\begin{figure*}[h!]
 \centering  \caption{SEE ATTACHED JPG FIGURE Plots of $R_{4570}$
against FWHM(H$\beta$) (left panel) and FWHM(H$\alpha$) (right
panel). Objects with detected FeII emission are denoted as filled
circles, and non-detections (3 $\sigma$ upper limits) are denoted as
crosses. } \label{f15}
\end{figure*}

\setcounter{figure}{15}
\begin{figure*}[h!]
 \centering  \caption{SEE ATTACHED JPG FIGURE Left panels: plots of
H$\beta$ luminosity versus the equivalent width of H$\beta$
(upper) and FeII4570 (lower). Right panels
show plots of
monochromatic nuclear luminosity versus the equivalent width of
H$\beta$ (upper) and FeII4570 (lower).} \label{f16}
\end{figure*}

\setcounter{figure}{16}
\begin{figure*}[h!]
 \centering  \caption{SEE ATTACHED JPG FIGURE Plots of the soft X-ray
photon index against the FWHM of H$\beta$ (left panel) and H$\alpha$
(right panel).} \label{f17}
\end{figure*}

\clearpage
\setcounter{figure}{17}
\begin{figure*}[h!]
 \centering  \caption{SEE ATTACHED JPG FIGURE The upper panel shows
the distribution of the $[OIII]\lambda5007/H\beta^{nc}$ ratio. Objects
with reliably determined values are shown in
the hallow histogram and
those with undetectable flux of H$\beta$ narrow component in
hatched ($[OIII]\lambda5007/H\beta^{nc}=10$ is assumed
for illustration purpose). Diagnostic diagram for the NLR of the
168 NLS1s with reliably measurements of all the 4 narrow emission
lines are plotted in the lower panel. The solid line
marks the
theoretical classification line between Seyfert 2s and starforming
galaxies (Kewley et al. 2000).} \label{f18}
\end{figure*}

\clearpage
\setcounter{figure}{18}
\begin{figure*}[h!]  
\caption{SEE ATTACHED JPG FIGURE Distribution of the relative
redshift for [OII] (a), [OIII] (b) and the broad components of
H$\beta$ (c) and H$\alpha$ (d). The shift of [OII] is relative to the
[NII] and [SII] doublets. The shift of [OIII] and the broad components
of Balmer emission lines
is relative to the low-ionization forbidden
lines, [OII], [NII], or [SII]. A few tens of ``blue outliers" are
pinpointed in the inset. } \label{f19}
\end{figure*}

\setcounter{figure}{19}
\begin{figure*}[h!]
\centering  \caption{SEE ATTACHED JPG FIGURE 
Sample spectra of
``blue outliers" (upper panel) and narrow line emitters (lower
panel). Vertical lines indicate the expected centroid of emission
lines according to
the low-ionization forbidden lines.} \label{f20}
\end{figure*}

\setcounter{figure}{20}
\begin{figure*}[h!]
 \centering  \caption{SEE ATTACHED JPG FIGURE Examples of
``UV-deficient" NLS1s. While most of these objects can be
explained by intrinsic extinction indicated by large Balmer
decrement (lower panel), the
rest cannot  (upper panel).} \label{f21}
\end{figure*}

\setcounter{figure}{21}
\begin{figure*}[h!]
 \centering  \caption{SEE ATTACHED JPG FIGURE An example of
``FeII-deficient" NLS1s. Note the very narrow Balmer emission
lines and also the very weak FeII multiplets.
The prominent HeII broad emission line ensures our
classifying it as a type~1 AGN.
} \label{f22}
\end{figure*}

\setcounter{figure}{22}
\begin{figure*}[h!]
 \centering  \caption{SEE ATTACHED JPG FIGURE Spectra of 6 NLS1s that
possibly show the MgII broad absorption trough. Three candidates, SDSS
J102021.21+121909.1, SDSS J115816.72+132624.2, and SDSS
J145724.01+452157.8, are selected from the SDSS DR4.} \label{f23}
\end{figure*}

\setcounter{figure}{23}
\begin{figure*}[h!]
 \centering
%\figurenum{26}
 \caption{SEE ATTACHED JPG FIGURE The relation between the stellar velocity
dispersion and the black hole mass for 308 NLS1s
for which
$\sigma_{*}$
is reliably measured. The dash line is the relation
given by Tremaine et al. (2002) for normal galaxies.} \label{f24}
\end{figure*}

\clearpage
\setcounter{figure}{24}
\begin{figure*}[h!]
\caption{SEE ATTACHED JPG FIGURE Montages of 33 NLS1s with
redshift less than 0.1, for which either the host galaxies are
viewed  face-on or the SDSS $3^{''}$ fiber aperture is dominated by
the bulge contribution.} \label{f25}
\end{figure*}

\clearpage

\setcounter{figure}{25}
\begin{figure*}[h!]
\centering  \caption{SEE ATTACHED JPG FIGURE Same as Figure 24 but for
33 nearby NLS1s.} \label{f26}
\end{figure*}

\clearpage

\setcounter{figure}{26}
\begin{figure*}[h!]
 \centering  \caption{SEE ATTACHED JPG FIGURE The correlation between
FWHM[NII] and $\sigma_{*}$ for 206 NLS1s with both parameters
reliably measured. The solid line denotes the
often used relation of $FWHM[NII]=2.35~\sigma_{*}$. } \label{f27}
\end{figure*}

\setcounter{figure}{27}
\begin{figure*}[h!]
 \centering  \caption{SEE ATTACHED JPG FIGURE Same as Figure \ref{f27}
but for $\sim 3~000$ Seyfert 2 galaxies with high S/N selected from
the SDSS. The dash line
marks the linear fit to the data.}
\label{f28}
\end{figure*}

\setcounter{figure}{28}
\begin{figure*}[h!]
 \centering
%\figurenum{26}
 \caption{SEE ATTACHED JPG FIGURE $M_{BH}-\sigma_{*}$ relation for 613
NLS1s in the sample. The dash line
marks the expected relation of
Tremaine et al. using $FWHM[NII]=2.35~\sigma_{*}$. } \label{f29}
\end{figure*}

\setcounter{figure}{29}
\begin{figure*}[h!]
 \centering
%\figurenum{28}
 \caption{SEE ATTACHED JPG FIGURE Spectra of 4 NLS1s with large amplitude
variability. Two epoch SDSS spectra are shown in red and green, and
the difference in blue. } \label{f30}
\end{figure*}

%%%%%%%%%%%%%%%%%%%%%%%%%%%%%%%%%%%%%%%%%%%%%%%%%%%%%%%%%%%%%%%%%%%

\end{document}